\documentclass[a4paper,11pt]{article}
\pdfoutput=1 
\usepackage{subcaption}
\usepackage{jcappub,aas_macros} 

\usepackage{graphicx}
\usepackage{dcolumn}
\usepackage{bm}
\usepackage{xcolor}
\usepackage{aas_macros}
\usepackage{hyperref}

\title{\boldmath Emulating Recombination with Neural Networks using Universal Differential Equations}

\author[a]{Ben Pennell}

\author[a,b,1]{Zack Li\note{Corresponding author.},}
\emailAdd{zackli@berkeley.edu}

\author[b,c]{James M. Sullivan}

\affiliation[a]{Canadian Institute for Theoretical Astrophysics, University of Toronto, Toronto, ON, Canada M5S 3H8}
\affiliation[b]{Berkeley Center for Cosmological Physics, University of California, Berkeley, CA 94720, USA}
\affiliation[c]{Department of Astronomy, University of California, Berkeley, CA 94720, USA}

\abstract{
With an aim towards modeling cosmologies beyond the $\Lambda$CDM paradigm, we demonstrate the automatic construction of recombination history emulators while enforcing a prior of causal dynamics. 
These methods are particularly useful in the current era of precision cosmology, where extremely constraining datasets provide insights into a cosmological model dominated by unknown contents. 
Cosmic Microwave Background (CMB) data in particular provide a clean glimpse into the interaction of dark matter, baryons, and radiation in the early Universe, but interpretation of this data requires knowledge of the Universe's ionization history.
The exploration of new physics with new CMB data will require fast and flexible calculation of this ionization history.
We develop a differentiable machine learning model for recombination physics using a neural network ordinary differential equation architecture (Universal Differential Equations, UDEs), building towards automatic dimensionality reduction and the avoidance of manual tuning based on cosmological model.
}

\begin{document}
\maketitle
\flushbottom

\section{Introduction}

Precise measurements of the Cosmic Microwave Background (CMB) have helped build a Standard Model of Cosmology. 
The current state of the art is dominated by data from the \emph{Planck} satellite \cite{Planck}, but increasingly precise data from current and planned experiments such such as from the Atacama Cosmology Telescope \citep[ACT,][]{thornton/etal:2016}, the South Pole Telescope \citep[SPT,][]{benson/etal:2014}, the Cosmology Large Angular Scale Surveyor \citep[CLASS,][]{class14}, the upcoming Simons Observatory \citep[SO,][]{so_forecast:2019}, and CMB-S4 \citep{cmbs4:2019} will shed further light on the cosmic mysteries of dark matter (DM) and dark energy (DE) implied by the Standard Model of Cosmology ($\Lambda$CDM).  

Calculation of the ionization history - the global mean ionization state of atomic species as a function of time - is a critical aspect of the interpretation of CMB data. 
Although the basic physics has been understood for more than half a century \cite{zeldovich68, Peebles}, the requirement of precise calculations of atomic physics has led to the development of a sequence of high-performance, physically-motivated emulators for the full ionization physics. 
The complexity began when the Universe sufficiently cooled, around redshift $z\approx1100$, when it was possible for free electrons to bind with nuclei to create neutral hydrogen and helium. The timescale of combining directly to the ground state is too large for a significant portion of neutral atoms to be formed that way. Instead, electrons are captured in ionized states and then drop to the ground state. For this reason, the full description of recombination is thousands of coupled ordinary differential equations, each corresponding to various ionization states of the electrons in the nuclei.

A number of high-quality emulators exist for recombination. \texttt{RECFAST} \cite{RECFAST} solves a modified 3-level atom approximation of recombination physics and uses well-chosen approximations and fudge factors to reach an accuracy suitable for the analysis of data from previous CMB experiments, but was already insufficient for the analysis of \emph{Planck} data. 
A more complete model of the global ionization history is provided by the \texttt{HYREC}  \cite{Hyrec} code.
\texttt{HYREC} resolves the out-of-equilibrium angular momentum substates not included in \texttt{RECFAST} and includes Lyman-$\alpha$-related radiative transfer effects.
\texttt{HYREC-2} \cite{Hyrec2} is a fast version of \texttt{HYREC} that employs several approximations but achieves a similar level of (high) accuracy compared to \texttt{RECFAST}.

These emulators involve a careful understanding of the physics to separate the aspects of the differential equations that contribute significantly to the dynamics and those aspects that can be safely neglected. In this work, we instead leverage recent developments in machine learning (ML) to learn these dynamics automatically, a step towards the automatic generation of physically motivated, interpretable emulators.  
Modern theory and computation power is capable solving the full set of differential equations, although with significant cost during inference. 
For this reason, an emulator would prove very useful for testing cosmological models and model extensions, and is an essential part of bridging the gap of being able to calculate recombination histories to using that knowledge to learn about cosmology itself. 

One of the primary goals of the modern cosmological program is to learn fundamental physics from astrophysical observations. 
Naturally, it is then desirable to develop model extensions of standard recombination physics, since upcoming datasets will be sensitive to such extensions.
This work provides an emulation method that separates the dependence on the background physics from the atomic physics of recombination.
Some changes to atomic physics, such as purely additive terms to the ODE, can be accommodated easily in our framework based on Universal Differential Equations (UDEs). 
As a by-product, we obtain an emulator for the ODE, and therefore, a differentiable forward model - obviating the need for manual tuning of approximate physics emulators like RECFAST.
This method explores the trade-off in flexibility and interpretability between emulators based on physical insight and the black-box emulation increasingly embraced by the cosmological community.

\section{Universal Differential Equations}
A system of coupled, first order ordinary differential equations describes the evolution of the state variables $\textbf{u}$ using their first derivatives $\dot{\textbf{u}}$.
These are functions of the state variables, denoted by $\textbf{u}$, input parameters, denoted by $\theta$ and a parameterization variable, often time, denoted by $t$.

\begin{equation}
\dot{\textbf{u}} = f(\textbf{u}, \theta, t)
\end{equation}

For recombination, the parameterization variable is redshift $z$. A Universal Differential Equation (UDE) is a neural network (NN) embedded into a differential equation solver.
The job of the UDE is to model this function,

\begin{equation}
f(\textbf{u}, \theta, z) = NN(\textbf{u}, \theta, z).
\end{equation}

From this the structure of UDEs becomes evident: the state variables, parameters, and redshift are taken as inputs. The first derivatives, $\dot{\textbf{u}}$, are the outputs.\cite{SciML},\citep[e.g., as applied in][]{ude_1,ude_2,ude_3} 

The value of UDEs is in not directly interpolating a measured time series, but rather to approximate the differential equations that govern them. 
Differential equation solvers are built to focus computation time on the challenging dynamics, by taking larger steps during regions with simpler, flatter dynamics. 
The differential equation solver that the NN is embedded into can then expedite training, especially with particularly challenging dynamics in the target time series. 
As a result, UDEs are a useful tool for approximating the physics of cosmological recombination.

\begin{figure}
    \centering
    \includegraphics[width=\columnwidth]{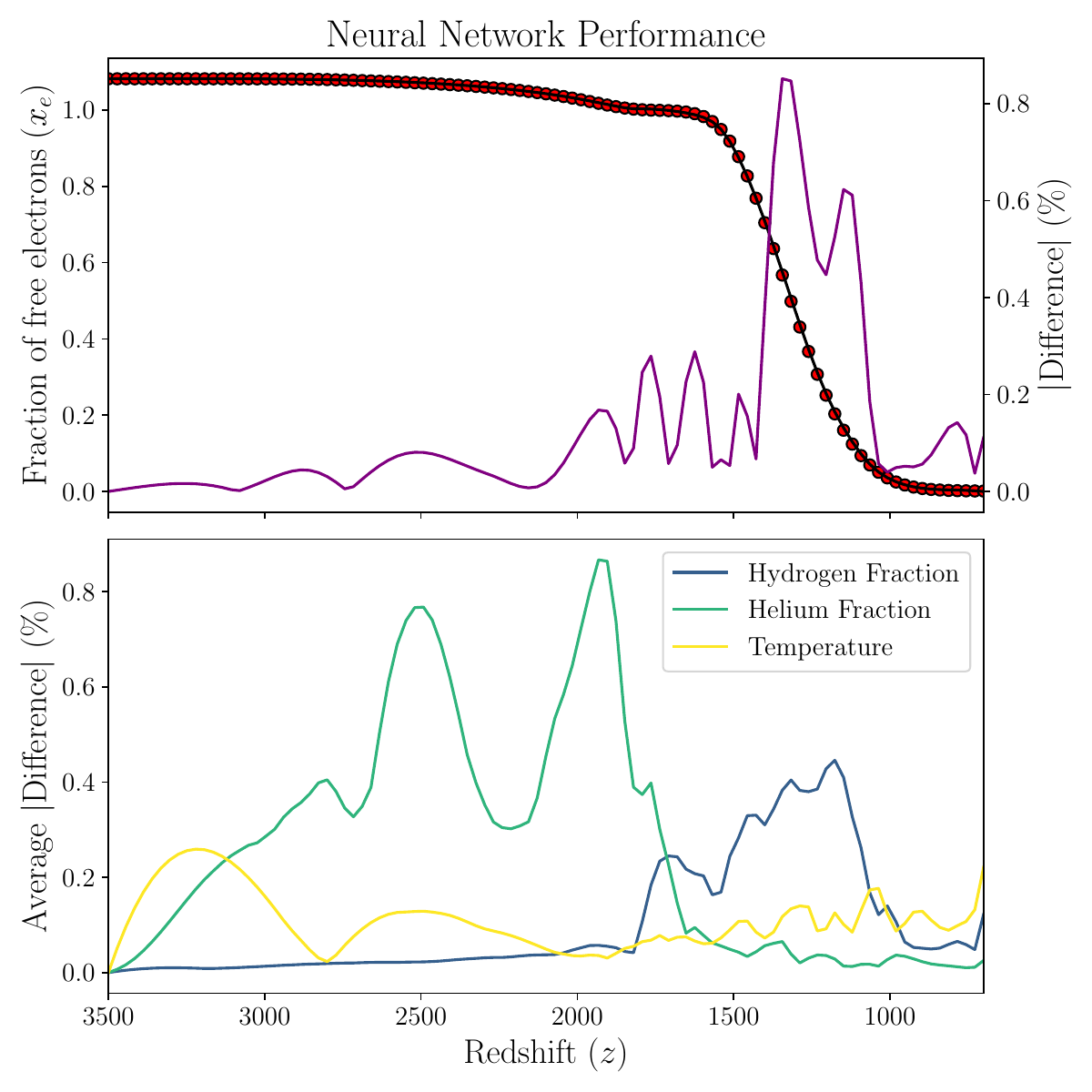}
    \caption{Demonstration of network fit by comparing the network output for cosmologies in the test set. In the top panel we compare the fraction of free electrons $x_e$ determined by the neural network in the solid black line and from \texttt{HYREC-2} with red circles. This is done using an ionization history using parameters that match the best fit values from Planck 2018 \cite{Planck}. The same panel also shows the percent difference between the network and training data in purple. The bottom panel shows the average percentage difference in network compared to training output across each test set ionization history.}
    \label{fig:NNHyrec}
\end{figure}

\section{Emulation}

The increasing cost of high-dimensional inference in data analyses makes fast surrogate models for physical processes highly desirable.
These so-called ``emulators'' reduce the runtime of physical forward models (such as ODE solvers) and furnish gradients of the approximate surrogate model - both aspects lead to faster inference.
Emulators have been highly successful in cosmology, in particular over the last decade, e.g. \cite{cosmicemu,auldnn,moped,cosmopower,pico,euclidemu2,arico_emu,bacco,cosmicnet,drew_field_emu,marco_emu,capse,kwan23_emu,boris_emu,moran_23_emu,boruah_emu,comet_emu,cosmicnet2,necola,matryoshka,kernel_emu,multifidelity_gp_emu,kokron_lagemu,hmcode2020,rsd_emu,wibking_emu,darkquest_emu,aemulus3,bird_emu,reion_emu,miratitan2,kwanemu1,pkann,coyoteext,coyote3}, and our goal here is to introduce a more physically-aware emulator of recombination at the level of the ODE.
Since this is the first time such a scheme has been used for emulation of the ionization history, we work with a relatively narrow scope in terms of the design space of cosmological parameters (around $\Lambda$CDM cosmology) as well as in dataset and network size.

We produced training data in the form of ionization histories with \texttt{HYREC-2}.
Each ionization history consisted of three time series: fraction of free Hydrogen nuclei, fraction of free Helium nuclei, and temperature. 
Each series was parameterized in terms of the redshift.
The varied parameters are the temperature of the CMB today $T_{\mathrm{CMB}}$, the baryon density $\Omega_{b}$, and the dark matter density $\Omega_{m}$.~\footnote{We note that varying all three of these parameters is not strictly necessary under standard assumptions, as we consider here, since the ionization history depends only on the ratios of each of the baryon and CDM density parameters to $T_{\rm cmb}^3$ \cite{ivanov_T0}. We thank Yacine Ali-Ha\"{i}moud for pointing this out.}
In total, we used forty-eight ionization histories with the three varied cosmological parameters sampled randomly in a latin hypercube within ten percent of Planck 2018 best-fit parameters to test the extent that a single UDE can learn several varied time series at once \cite{Planck}.

Each sample had a hundred evenly spaced points in redshift between $z=3500$ and $z=700$, the epoch with the most complicated dynamics in the ionization history \cite{Peebles, zeldovich68}.
The two series corresponding to hydrogen and helium can be combined into a single value corresponding to the fraction of free electrons, $x_e$. This is more interpretable for the final answer, and so is how we compare the output from the network to the test set in Figure~\ref{fig:NNHyrec}.

\begin{align}
    f_{\mathrm{He}} &= \frac{Y_p}{3.9715(1 - Y_p)} \\
    x_{e} &= x_{\mathrm{H}} + f_{\mathrm{He}} x_{\mathrm{He}}
\label{eqn:fHe}
\end{align}

We tested several NN architecture choices. 
Using four hidden layers, we found that thirty neurons per layer was able to learn the dynamics to within 1\% error on average.
A hyperbolic tangent activation function was used in each hidden layer. 
The network takes in seven inputs and has three outputs. 
The inputs are: the three state variables, the redshift, and the values of the varied cosmological parameters $\mathbf{\theta} = (\Omega_{b},\Omega_{m}, T_{\mathrm{CMB}})$. 

\begin{equation}
   \dot{\mathbf{x}}(z)
    = NN(\mathbf{x}(z), p, \mathbf{\theta}, z)
\end{equation}

Where the internal ODE ionization history is $\mathbf{x}(z) = \left( x_{\mathrm{H}}(z),x_{\mathrm{He}}(z),T(z) \right)$, and $p$ are the NN weights and biases.
We apply a normalization procedure as described in Section~\ref{subsec:Normalization}.

The outputs are the derivatives of each of the state variables. 
A Runge-Kutta differential equation solver of order five \cite{Tsit5} is then solved over the same redshift interval as the training data, using the NN as the system of equations. 
We used the $L2$-regularized (weight decay) $\ell_{1}$ loss to train the network and find NN parameters $p$:
\begin{equation}
L(p) = \sum_{i} \left|  \mathbf{y}(z_{i}) - \int_{z_{\mathrm{max}}}^{z_{i}}  NN(\mathbf{x}(z), p, \mathbf{\theta},z) \, dz  \right| + \lambda p^{2}
\label{eqn:loss}
\end{equation}
where the training data is $\mathbf{y}(z)$, and the second regularization term is described in Section~\ref{subsec:WeightDecay}.

\subsection{Training} \label{subsec:training}
The NN is pre-trained by estimating the derivative of each time series by calculating the slope at each point, and the NN is trained directly on this approximation for 600 iterations of gradient descent. 
Then, the network is trained using the ADAM gradient descent optimizer\cite{ADAM} and converged within 5000 iterations.
To mitigate finding a sub-optimal minimum of the loss, regularization using a training schedule with data batching (Section~\ref{subsec:Training}), weight decay (Section~\ref{subsec:WeightDecay}), and sampling several different random initialization of the network parameters. In total, we trained the network with 150 random parameter initializations, using the same network architecture, and used the best-performing parameters for the network predictions used in the results of this paper.
Further regularization techniques can be applied to improve fitness and training speed, which will be left for future work.

\begin{figure}
    \centering
    \includegraphics[width=\columnwidth]{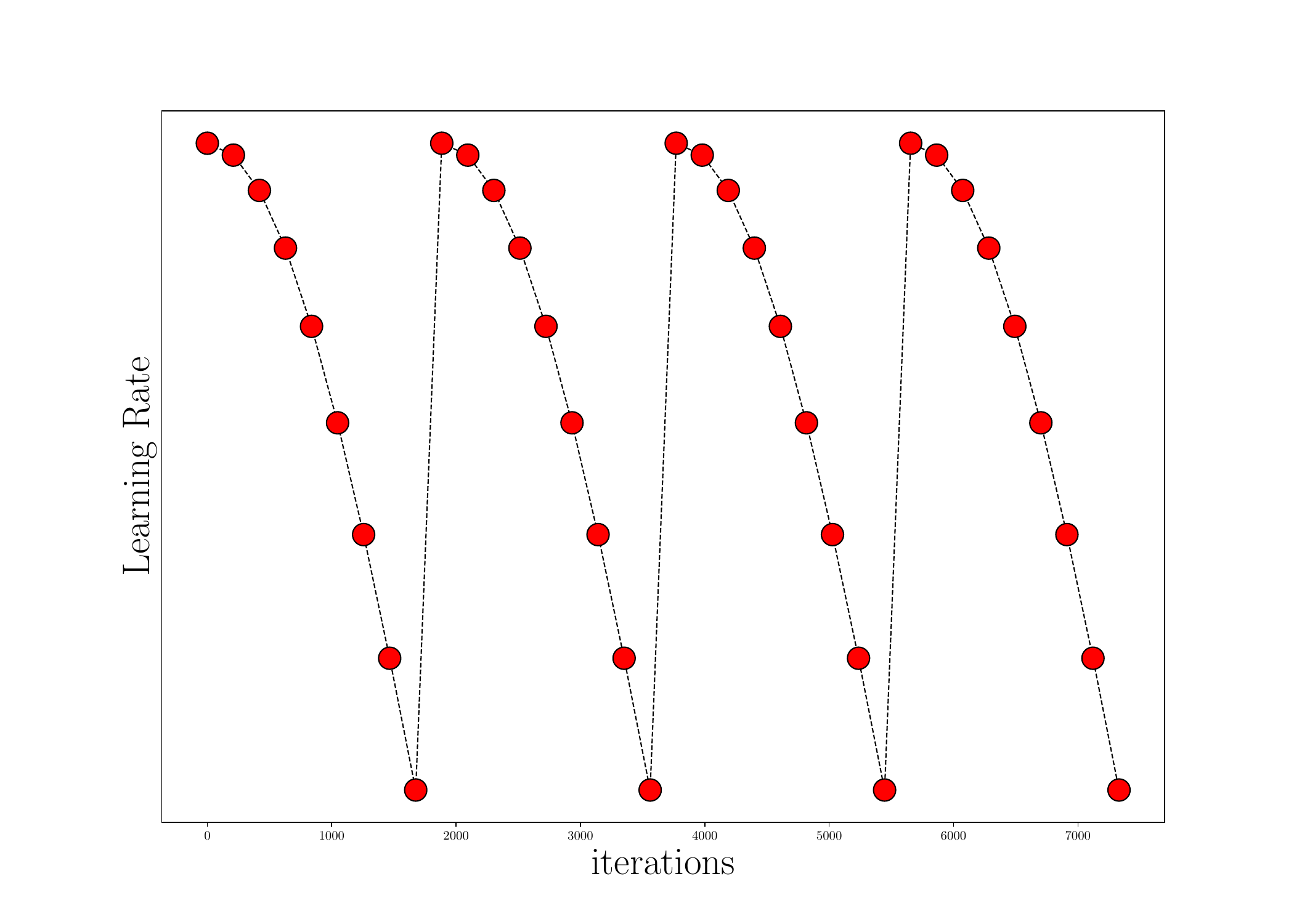}
    \caption{Example of a cosine learning schedule, which is used in training the neural network which is a repeated first quadrant of a cosine curve.}
    \label{fig:Schedule}
\end{figure} 

\subsection{Training Schedule} \label{subsec:Training}
 The training dependence on the training schedule should be noted. In gradient descent optimization, the size of the steps taken down the slope on parameter space changes how well a NN learns. A sufficiently small step will take longer to converge on a solution, will converge onto local minima, and will eventually reach the best attainable answer in its domain of parameter space. A larger step size can move past local minima, and towards a solution faster, but will not be able to reach the lowest point in its domain in parameter space.  For UDEs, a poor choice of learning rate makes the NN unable to train, as a result of the stiffness of the differential equations. 
 
A mix of learning rates must be employed. We found that an initial learning rate of 0.01 was the smallest that could reach an acceptable error. Progressively smaller learning rates would then be subsequently used to locate better solutions. After reaching the smallest learning rate, the process would repeat. The learning rates were chosen by a cosine curve, so that most of the learning time was spent with a high learning rate, then fewer iterations at a lower learning rate, an example can be seen in figure \ref{fig:Schedule}. The network had three hundred iterations per learning rate, and sampled thirty learning rates over three cosine waves.

\begin{figure}
    \centering
    \includegraphics[width=\columnwidth]{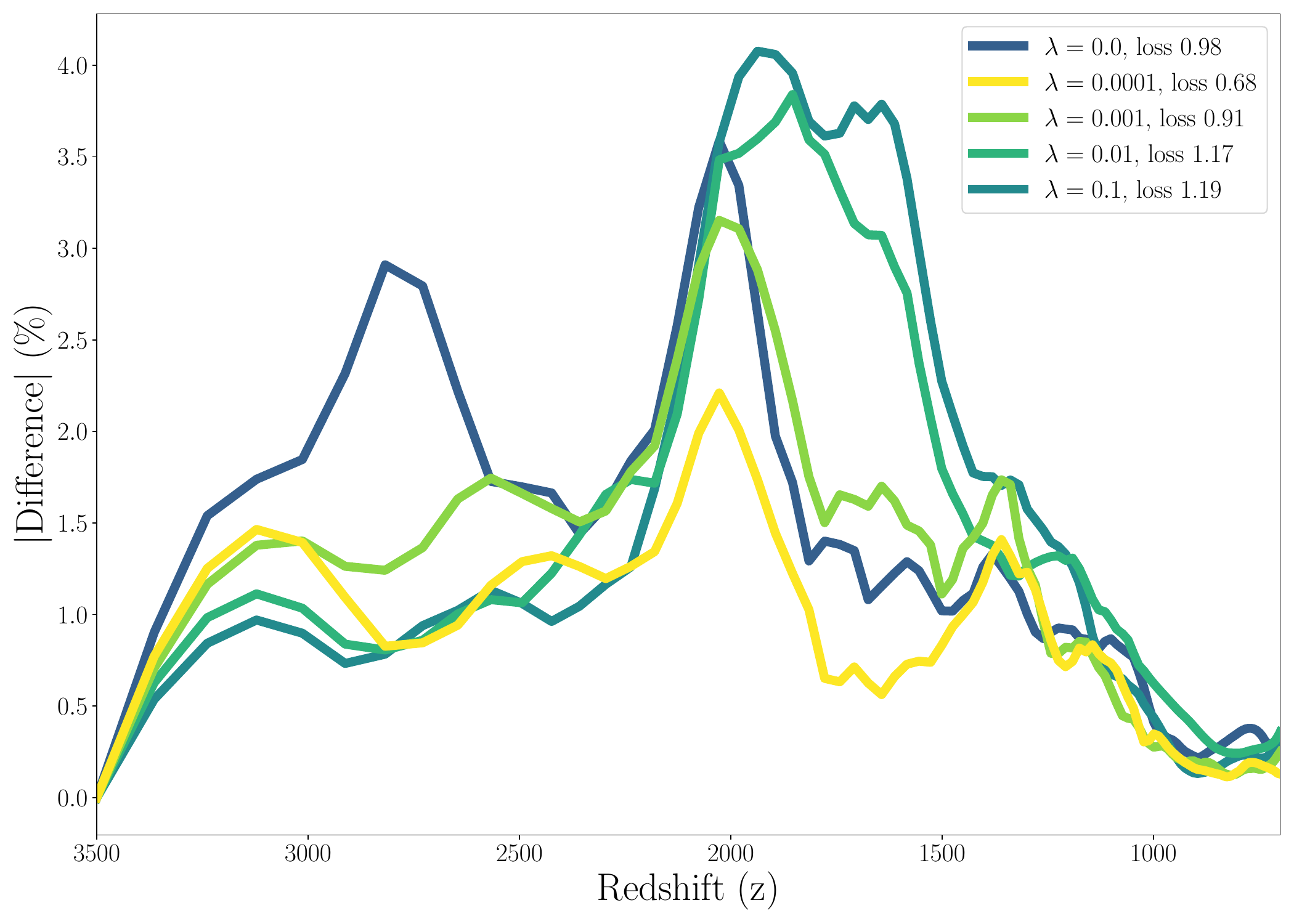}
    \caption{Difference between network output and training data plotted for example neural network with the same architecture as the emulator with different values of weight decay coefficient. A coefficient of zero means that weight decay is not being used. Coefficients with values smaller than 0.0001 made little change compared to no weight decay.}
    \label{fig:WeightDecay}
\end{figure}

As an additional way to increase the generality of the NN over the input parameter space, the training data was batched. Of the forty eight training ionization histories, at each iteration only sixteen of them at random would be used to train the NN. This both prevents the NN from over-fitting to training data and to better explore parameter space.

\subsection{Weight decay} \label{subsec:WeightDecay}
To combat the NN getting trapped in a particular region of parameter space, weight decay is employed. Weight decay adds the squared norm of all the parameters of the NN, multiplied by a constant, to the loss, which can be seen in Equation~\ref{eqn:loss}. 
This punishes large weights and thus prevents the NN from converging on certain parameters being so large that gradient descent cannot improve the fitness. Several coefficients were tested; too large of a coefficient makes the network unable to learn the dynamics, while too small of a coefficient makes little change. As can be seen in Figure~\ref{fig:WeightDecay}, large coefficients do not improve performance, but a sufficiently small value does. A coefficient of $0.0001$ decreased the loss by a factor of almost a third overall, and made the network output smoother.

\subsection{Normalization} \label{subsec:Normalization}
A challenge in training is when the output is expected to span multiple orders of magnitude. By nature of the hyperbolic tangent being used as the activation function between layers, the last hidden layer is constrained to within positive and negative one, which limits the range of possible outputs to the order of unity, while the expected output is on the order of ten thousand. To combat this, we normalized the network inputs and scaled up the network outputs when comparing to the training data to determine network fitness.

\section{Results \& Discussion}
We illustrate the generalization of the NN to multiple input parameters by using three varied cosmological parameter combinations in the training and test sets. The accuracy of the emulator on the test set can be seen in the bottom panel of Figure~\ref{fig:NNHyrec}. The average difference between the NN output and test set is 0.16\%.
Among these, the test set included the Planck 2018 best fits. The network's fitness on the Planck 2018 best fits can be seen in Figure~\ref{fig:NNHyrec}.

This result is both a first step towards autonomous complete emulation of recombination physics, and an illustration of the efficacy of UDEs as a tool for creating physics-informed emulators. 
The emulator has sampled three parameters and has sub-percent accuracy inferring within the sampled range, as can be seen in the bottom panel of Figure~\ref{fig:NNHyrec}, which provides an efficient method for calculating ionization histories for other cosmologies in the range. Sub-percent error makes this network a comparable emulator of the whole physics captured in \texttt{HYREC-2} as the approximations in \texttt{RECFAST}.
Neural networks are universal function approximators, and so its approximation of the differential equations of recombination is only limited by available computational resources.
Batching of the ionization histories and training on different elements of the dynamics separately would likely help increase the effectiveness of training \citep[i.e.][]{diffeqfluxjl, multiple_shooting}, and is left for future work to investigate.

In this paper, the emulator was trained on cosmologies in the $\Lambda$CDM paradigm, to match with the approximations in \texttt{HYREC-2}. We have demonstrated that UDEs can be used to automatically tune parameters to reduce the dimensionality of the problem, in lieu of manually determined approximations such as what was done in \texttt{RECFAST}.
In future work, the emulator could be trained with more input parameters varied over wider ranges to tackle different and more complicated problems. 
With the use of only 48 ionization histories in the training set sampled narrowly around the \emph{Planck} best fits, our emulator is only relevant to a small number of cosmological models. Future work is needed to create a more general emulator with a broader scope.
There is no reason to suspect that increasing the number of input parameters would significantly hinder training, especially since the dependence of ionization physics on cosmological parameters is typically very smooth.

In this paper, we compressed a large number of coupled ODEs into a system with three variables, automatically producing an emulator like RECFAST.  
However, the physics of recombination is independent of cosmology for a given state of the Universe, and in principle, we could eventually train a more general emulator that does not require cosmology as inputs like RECFAST (aside from $\Omega_b$), but purely learns the dynamics of the atomic physics given $H(z)$. 
This would largely prevent the need for retraining under different cosmologies, and would accelerate model building. 
We leave this for future work.

\appendix
Our code is hosted publicly on GitHub at the following link: \url{https://github.com/BenPennell/recombination-emulator}

\acknowledgments
BP acknowledges support from CITA over the course of this
research, including through the CITA Summer Undergraduate Re-
search Fellowship and an NSERC Undergraduate Student Research
Award.
ZL acknowledges support by the Natural Sciences and Engineering Research Council of Canada (NSERC), funding references CITA
490888-16 and RGPIN-2020-03885.
JMS was partially supported by the U.S. Department of Energy, Office of Science,
Office of Workforce Development for Teachers and Scientists, Office of Science Graduate Student Research (SCGSR) program. 
The SCGSR program is administered by the Oak Ridge Institute for Science and Education (ORISE) for the DOE. ORISE is managed by ORAU under contract number DE-SC0014664.
We thank Yacine Ali-Ha\"{i}moud for comments on an earlier version of this draft.

\nocite{*}
\bibliographystyle{JHEP}
\bibliography{apssamp}

\end{document}